\documentclass[nobibnotes,preprintnumbers,aps,prl,superscriptaddress,twocolumn,twoside,sort&compress, longbibliography]{revtex4-1}				 
\usepackage{graphicx}																												 
\usepackage{titlesec}	
\usepackage{multirow}	
\usepackage{gensymb}		
\usepackage{balance}
						
\usepackage{blindtext} 
\usepackage[colorlinks=true,linkcolor=blue,citecolor=red,urlcolor=blue]{hyperref}													 

\usepackage{booktabs}
\usepackage{multirow}

\usepackage{tablefootnote}

\begin{document}																													 
%

\newcommand{\kvec}{\mbox{{\scriptsize {\bf k}}}}
\newcommand{\lvec}{\mbox{{\scriptsize {\bf l}}}}
\newcommand{\qvec}{\mbox{{\scriptsize {\bf q}}}}

\newcommand{\mycomment}[1]{}
\def\eq#1{(\ref{#1})}
\def\fig#1{\hspace{1mm}Fig. \ref{#1}}
\def\figur#1{\hspace{1mm}Figure \ref{#1}}
\def\tab#1{\hspace{1mm}Table \ref{#1}}
\title{
Theoretical prediction of strong-coupling superconductivity in a hypothetical NaAlH$_3$ phase at ambient pressure}

\author{Izabela A. Wrona}
\affiliation{Faculty of Electrical Engineering, Czestochowa University of Technology, Armii Krajowej 17, Czestochowa, 42-200, Poland
}
\author{Yinwei Li}
\affiliation{Laboratory of Quantum Materials Design and Application, School of Physics and Electronic Engineering, Jiangsu Normal University, Xuzhou 221116, China}
\author{Radoslaw Szczesniak}
\affiliation{Department of Physics, Czestochowa University of Technology, Ave. Armii Krajowej 19, 42-200 Czestochowa, Poland}
\author{Artur P. Durajski} \email{artur.durajski@pcz.pl}
\affiliation{Faculty of Electrical Engineering, Czestochowa University of Technology, Ave. Armii Krajowej 17, Czestochowa, 42-200, Poland
}
\date{\today} 

\begin{abstract}
We present a comprehensive first-principles investigation of a hypothetical cubic $Pm\overline{3}m$ phase of the ternary hydride NaAlH$_3$, focusing on its lattice dynamics, electronic structure, and electron–phonon-mediated superconducting properties at ambient pressure. Using density functional theory and the Migdal-Eliashberg formalism, we find an exceptionally strong electron-phonon coupling ($\lambda=2.23$), resulting in a superconducting critical temperature of up to $73.7$ K for a Coulomb pseudopotential $\mu^* = 0.1$. Phonon dispersion calculations, complemented by \textit{ab initio} molecular dynamics simulations, indicate dynamic and thermal stability within the adopted theoretical framework. The electronic structure exhibits a metallic character with substantial contributions from Al- and Na-derived states at the Fermi level. The resulting superconducting gap ratio ($2\Delta(0)/k_B T_c \approx 4.8$) and specific heat jump ($\Delta C/\gamma T_c \approx 2.2$) significantly exceed BCS weak-coupling predictions, highlighting the strong-coupling nature of superconductivity in this hypothetical phase.
\end{abstract}
\maketitle
%

\section{I. Introduction}
The pursuit of high-temperature superconductivity in hydrogen-rich materials has been a central focus of condensed matter physics in recent decades. Following the seminal works of Ashcroft \cite{ Ashcroft1968A, Ashcroft2004A}, which proposed that metallic hydrogen and hydrogen-dominant compounds could exhibit high critical temperatures ($T_c$), extensive theoretical and experimental efforts have been devoted to exploring binary hydrides under high pressure. This has led to the discovery of remarkable superconductors such as H$_3$S ($T_c = 203$ K at $155$ GPa) \cite{li2014, drozdov2015,einaga2016, PhysRevB.105.L220502}, LaH$_{10}$ ($T_c = 250-260$ K at $170-190$ GPa) \cite{drozdov2019,somayazulu2019}, CaH$_6$ ($T_c = 210$ K at $160$ GPa) \cite{salke2019}, YH$_6$ ($T_c = 224$ K at $166$ GPa) \cite{troyan2021} and YH$_9$ ($T_c = 243$ K at $201$ GPa) \cite{kong2021}. However, the extreme pressures required for their stabilization present significant challenges for practical applications.

Recent advances in materials discovery have highlighted ternary hydrides as a promising class of high-temperature superconductors, offering greater compositional flexibility and enhanced properties at reduced pressures compared to binary systems \cite{PhysRevMaterials.7.L101801, Durajski2021, doi:10.1021/acs.jpcc.3c07213, PhysRevB.109.014501}. Particularly noteworthy are complex hydrides incorporating light elements (e.g., Al, B, Be, C), which exhibit exceptional phonon-mediated superconductivity due to their favorable electronic structures and high-frequency hydrogen-derived phonon modes \cite{ALAM2024107498, Zhang2022LaBeH8, wrona2024}. 
Among these systems, the La-B-H and La-Be-H have demonstrated remarkable superconducting behavior. Theoretical studies first identified the $Fm\overline{3}m$ structure for the La-B-H system, predicting $T_c$ values of $126-156$ K at $50-55$ GPa \cite{Di_Cataldo2021, liang2021}. This structural motif was subsequently extended to La-Be-H compounds, with first-principles calculations suggesting a $T_c \sim 185$ K for $Fm\overline{3}m$-LaBeH$_8$ above 98 GPa, maintaining dynamical stability down to 20 GPa \cite{Zhang2022LaBeH8}. Experimental validation followed in 2023 through high-pressure synthesis in diamond anvil cells, where X-ray diffraction and transport measurements confirmed superconductivity in LaBeH$_8$ with $T_c = 110$ K at $80$ GPa \cite{Song_2023}. 
Further progress emerged with the discovery of LaB$_2$H$_8$, which adopts a tetragonal $I4/mmm$ structure stabilized by strong B-H covalent bonding. This system exhibits a lower stabilization pressure ($59$ GPa) while maintaining high superconducting performance ($T_c > 105$ K at $90$ GPa) \cite{song2024A}, underscoring the critical role of structural chemistry in optimizing high-$T_c$ behavior at reduced pressures.

Notably, alanates (a class of materials containing [AlH$_4$]$^-$ or [AlH$_6$]$^{3-}$ complexes) have demonstrated intriguing superconducting behavior in theoretical studies \cite{Di_Cataldo_2023}. For example, a recent theoretical study on calcium alanates (Ca-Al-H) predicts that CaAlH$_7$ becomes dynamically stable at $50$ GPa and exhibits a superconducting critical temperature of $60$ K \cite{Di_Cataldo_2023}. RbAlH$_3$ remains stable under near ambient pressure ($15$ GPa), and may exhibit high-temperature superconductivity ($T_c=86$ K) \cite{Mingyang_2024}. Even lower pressures were obtained for KAlH$_3$ ($0-7$ GPa) with $T_c$ equals $52-96$ K \cite{Mingyang_2024, Wines_2024}.
This finding suggests that certain alanates can display superconducting properties driven by strong electron-phonon coupling when subjected to moderate pressures.

In this context, sodium aluminum hydride (NaAlH$_3$) presents a compelling case for investigation. As a member of the alanate family, it combines light constituent elements with a hydrogen-rich stoichiometry, both of which are key indicators of potential high-$T_c$ superconductivity \cite{WRONA2024101499}. Moreover, NaAlH$_3$ has been extensively studied for its hydrogen storage capabilities, with research focusing on its structural stability under ambient conditions \cite{KHAN2024596}. 
However, there have been no theoretical or experimental reports on the potential superconductivity of NaAlH$_3$, and its possible superconducting properties remain unexplored. Investigating NaAlH$_3$ in this new context could reveal novel insights into its electronic behavior and contribute to the broader search for high-temperature superconductors at ambient pressure.

This work presents a comprehensive first-principles investigation of a hypothetical NaAlH$_3$ phase, focusing on its electronic structure, lattice dynamics, and electron–phonon-mediated superconducting properties at ambient pressure. We address three key questions within the adopted theoretical framework: (i) whether such a phase can exhibit dynamic stability at ambient pressure, (ii) which electronic and phononic features are responsible for strong electron–phonon coupling, and (iii) how its predicted superconducting behavior compares with that of established high-pressure hydrides. Our analysis of the electronic density of states, phonon dispersion, and electron–phonon coupling reveals exceptionally strong coupling ($\lambda = 2.23$), leading to a high superconducting critical temperature of up to $73.7$ K for a reasonable choice of the Coulomb pseudopotential. These properties originate from the interplay between Al–H covalent bonding and Na-derived metallic states, which enhances phonon-mediated pairing.

\section{II. Methods}
\subsection{First-principles calculations}
The electronic and phononic properties of NaAlH$_3$ were investigated using first-principles calculations based on density functional theory (DFT), as implemented in the Quantum ESPRESSO package \cite{QE, QE2}. The Perdew-Burke-Ernzerhof functional revised for solids (PBEsol) \cite{PBE} was employed to accurately describe the exchange-correlation interactions, in conjunction with ultrasoft pseudopotentials \cite{GARRITY2014446, GBRV}. A plane-wave basis set was used with a kinetic energy cutoff of $40$ Ry for wavefunctions and $430$ Ry for charge density, ensuring well-converged total energies and forces. Structural relaxations were performed using the Broyden-Fletcher-Goldfarb-Shanno (BFGS) quasi-Newton algorithm \cite{bfgs}, with convergence criteria set to $10^{-10}$ Ry for total energy, $10^{-10}$ Ry/Bohr for atomic forces, and $10^{-10}$ kbar for pressure. To achieve high accuracy in electronic structure calculations, we employed a dense Monkhorst-Pack \cite{monkhorst1976} $k$-point mesh of $24\times24\times24$ for Brillouin zone sampling. A Gaussian smearing method with a width of $0.01$ Ry ($\sim$0.136 eV) was used throughout the self-consistent field (SCF) calculations to ensure convergence in the metallic state.
For the evaluation of the electronic density of states (DOS), a significantly denser $84\times84\times84$ $k$-point mesh was employed to enhance resolution. The DOS was likewise calculated using the Gaussian smearing method with a broadening parameter of $0.01$ Ry, ensuring consistency with the SCF calculations. This approach enables reliable representation of metallic features near the Fermi level while maintaining numerical stability.
Phonon and electron–phonon coupling properties were computed using density functional perturbation theory (DFPT) \cite{Baroni}, employing a $6 \times 6 \times 6$ $q$-point grid. The phonon dispersion and vibrational modes were verified to be converged with respect to this grid. For the evaluation of electron–phonon matrix elements and the computation of the Eliashberg spectral function $\alpha^2F(\omega)$ and the cumulative coupling parameter $\lambda(\omega)$, we used the same $24 \times 24 \times 24$ Monkhorst-Pack $k$-point mesh as in the SCF calculations. A Gaussian broadening of $0.005$~Ry was applied in the double-delta integration required for $\alpha^2F(\omega)$. This smearing width was determined through convergence testing and ensures a robust balance between spectral resolution and numerical stability.
To further evaluate the dynamical stability of NaAlH$_3$ at a temperature close to $T_c$, we performed \textit{ab initio} molecular dynamics (AIMD) simulations using CP2K software \cite{cp2k}. 

\subsection{Migdal-Eliashberg approach}
Conventional superconductivity arises from the formation of Cooper pairs, in which an effective attractive interaction between electrons, mediated by electron-phonon coupling, overcomes the inherent Coulomb repulsion between them. Given the strong electron-phonon coupling strength ($\lambda > 2$) in NaAlH$_3$, we employed the Migdal-Eliashberg (ME) formalism to describe its superconducting properties accurately. A central quantity in the ME theory is the Eliashberg spectral function, $\alpha^2F(\omega)$ defined as:
\begin{equation}
\alpha^2F(\omega) = {1\over 2\pi N(\varepsilon_F)}\sum_{{\bf q}\nu}
                    \delta(\omega-\omega_{{\bf q}\nu})
                    {\gamma_{{\bf q}\nu}\over\hbar\omega_{{\bf q}\nu}},
\end{equation}
where the phonon linewidth $\gamma_{\mathbf{q}\nu}$ is:
\begin{eqnarray}
\gamma_{{\bf q}\nu} &=& 2\pi\omega_{{\bf q}\nu} \sum_{ij}
                \int {d^3k\over \Omega_{BZ}}  |g_{{\bf q}\nu}({\bf k},i,j)|^2
                    \delta(\varepsilon_{{\bf q},i} - \varepsilon_F) \\\nonumber  && \times\delta(\varepsilon_{{\bf k}+{\bf q},j} - \varepsilon_F).
\end{eqnarray}
Here, $N(\varepsilon_F)$ represents the density of states at the Fermi level $\varepsilon_F$ and $g_{{\bf q}\nu}({\bf k},i,j)$ denotes the electron-phonon matrix element.

The isotropic Migdal-Eliashberg equations for the superconducting order parameter function $\varphi_{n}=\varphi\left(i\omega_{n}\right)$ and the electron mass renormalization function $Z_{n}= Z\left(i\omega_{n}\right)$ take the following form \cite{Eliashberg1960A, Marsiglio1988A, MARSIGLIO2020168102}:
\begin{equation}
\label{r1}
\varphi_{n}=\frac{\pi}{\beta}\sum_{m=-M}^{M}
\frac{\lambda_{n,m}-\mu^{\ast}\theta\left(\omega_{c}-|\omega_{m}|\right)}
{\sqrt{\omega_m^2Z^{2}_{m}+\varphi^{2}_{m}}}\varphi_{m},
\end{equation}
\begin{equation}
\label{r2}
Z_{n}=1+\frac{1}{\omega_{n}}\frac{\pi}{\beta}\sum_{m=-M}^{M}
\frac{\lambda_{n,m}}{\sqrt{\omega_m^2Z^{2}_{m}+\varphi^{2}_{m}}}
\omega_{m}Z_{m},
\end{equation}
where the electron-phonon interaction pairing kernel is defined as:
\begin{equation}
\label{r3}
\lambda_{n,m}= 2\int_0^{\infty}d\omega\frac{\omega}
{\left(\omega_n-\omega_m\right)^2+\omega ^2}\alpha^{2}F\left(\omega\right).
\end{equation}
Moreover, $\mu^{\ast}\in\{0.1, 0.15, 0.2\}$ denotes the Coulomb pseudopotential, which accounts for the depairing interaction between electrons. The superconducting order parameter is given by $\Delta_n=\varphi_{n}/Z_n$, and the superconducting transition temperature $T_c$ is determined from the condition $\Delta_{n=1}(\mu^{\ast}, T=T_c)=0$. The Heaviside step function $\theta$ enforces a cutoff frequency, set to ten times the maximum phonon frequency. The inverse temperature is defined as $\beta=1/k_BT$, where $k_B$ is the Boltzmann constant. 

The Eliashberg equations were solved iteratively using an in-house developed code. The convergence criterion required the maximum error tolerance between successive iterations to be below $10^{-10}$. To ensure numerical stability, we employed a sufficiently large number of Matsubara frequencies, $\omega_{n}=\left(\pi/\beta\right)\left(2n-1\right)$, with $n=0,\pm 1,\pm 2,\dots,\pm M$ and $M=1100$ \cite{Szczesniak2006A, Durajski2020, Durajski2021, Durajski2023}.

After solving the Eliashberg equations, the free energy difference between the superconducting and normal states, $\Delta F$, was computed using the relation \cite{Carbotte1990A}:
\begin{eqnarray}
\label{rF}
{\Delta F}&=&-{2\pi}T{N(\varepsilon_F)}\sum_{n=1}^{M}
\left(\sqrt{\omega^{2}_{n}+\Delta^{2}_{n}}- \left|\omega_{n}\right|\right)\\ \nonumber
&\times&(Z^{S}_{n}-Z^{N}_{n}\frac{\left|\omega_{n}\right|}
{\sqrt{\omega^{2}_{n}+\Delta^{2}_{n}}}),
\end{eqnarray}
where $Z^{S}_{n}$ and $Z^{N}_{n}$ represent the wave function renormalization factors for the superconducting ($\Delta_{n=1}\neq 0$) and normal ($\Delta_{n=1}=0$) states, respectively.

The specific heat difference between the superconducting and normal states, $\Delta C=C^{S}-C^{N}$, was evaluated as:
\begin{equation}
\label{rC}
{\Delta C}=-T\frac{d^{2}\Delta F}{dT^{2}},
\end{equation}
where the specific heat in the normal state was calculated using the expression $C^{N}=\gamma T$ and the Sommerfeld constant $\gamma$ is given by:
\begin{equation}
\gamma= \frac{2}{3} \pi^{2} \left(1+\lambda\right) k_{B}^2 N(\varepsilon_F).
\end{equation}

\section{III. Results and discussion}

\begin{figure}[!t]
    \centering
    \includegraphics[width=1\linewidth]{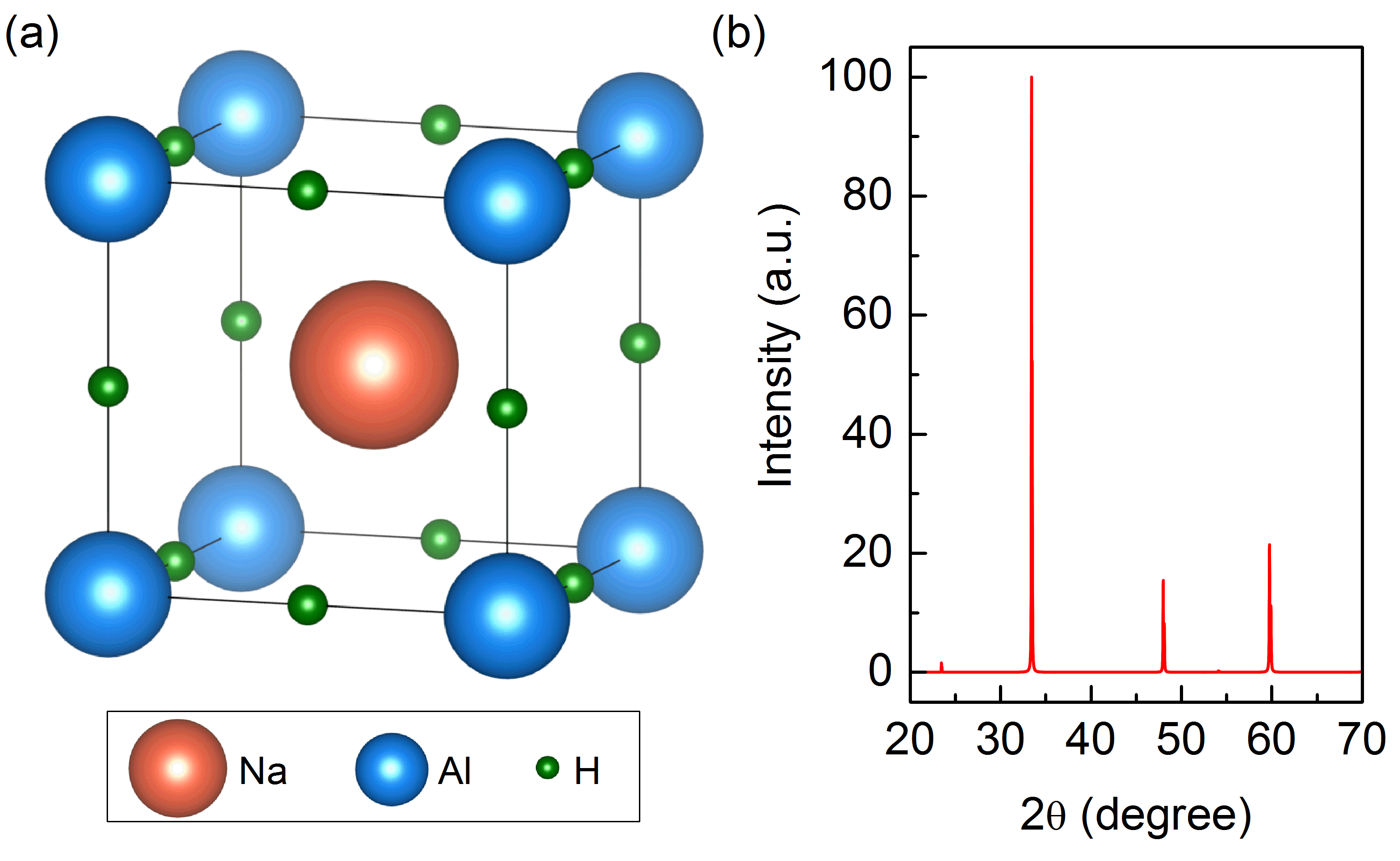}
    \caption{(a) Optimized crystal structure of NaAlH$_3$ in the cubic Pm$\bar{3}$m phase, obtained from fully relaxed variable-cell DFT calculations at $0$ GPa. (b) Simulated X-ray diffraction (XRD) pattern based on the relaxed equilibrium structure, confirming the phase purity and high symmetry of the compound.}
    \label{fig01}
\end{figure}

NaAlH$_3$ is modeled in the cubic space group $Pm\overline{3}m$ (No.~221), with Na, Al, and H atoms occupying the Wyckoff positions $1b$ $(0.5, 0.5, 0.5)$, $1a$ $(0, 0, 0)$, and $3d$ $(0, 0, 0.5)$, respectively. First-principles calculations yield an equilibrium lattice parameter of $3.789$~\AA\ at ambient pressure, consistent with previous theoretical reports~\cite{XU2024434, KHAN2024596}. The computed lattice constants ($a_0$), volumes ($V$), and Al–H and Na–H bond lengths characterizing the NaAlH$3$ structure under ambient (0~GPa) and near-ambient (0.1 and 0.2~GPa) conditions are summarized in \tab{tab01}. The minimal bond-length changes ($\Delta{\rm Al-H} < 0.003$~\AA) under pressure suggest a rigid covalent framework. As illustrated in \fig{fig01}, the optimized cubic structure of NaAlH$_3$ exhibits a well-ordered, periodic atomic arrangement. Simulated X-ray diffraction (XRD) patterns using Cu-K$\alpha$ radiation with a wavelength of $1.5401$~{\AA} are consistent with the assumed $Pm\overline{3}m$ symmetry, supporting the internal structural consistency of the model.

To assess the relative thermodynamic stability of the hypothetical $Pm\overline{3}m$ phase of NaAlH$_3$ at ambient pressure, we evaluated its formation energy ($\Delta E_f$) with respect to the elemental reference states according to
\begin{equation}
\Delta E_f = E{\text{tot}} - \left(E_{\text{Na}} + E_{\text{Al}} + 3E_{\text{H}}/4 \right).
\end{equation}
Here, $E_{\text{tot}}$ denotes the total energy of NaAlH$_3$, while $E{\text{Na}}$, $E_{\text{Al}}$, and $E_{\text{H}}$ correspond to the ground-state energies of bulk bcc Na (space group $Im\overline{3}m$, No.~229), bulk fcc Al (space group $Fm\overline{3}m$, No.~225), and crystalline hydrogen in its ground-state structure from the HEX database (space group $P6_3/mmc$, No.~194)~\cite{Giannessi2024}, respectively.

The calculated formation energy, $\Delta E_f = 0.22$~meV/atom, indicates that the considered NaAlH$_3$ phase is thermodynamically metastable with respect to decomposition into the elemental constituents, but with a formation energy that is only marginally positive. Given the small magnitude of $\Delta E_f$, this result may be sensitive to contributions beyond static DFT, such as zero-point motion and finite-temperature entropic effects. We note that this formation energy only reflects stability relative to the elemental phases and does not account for competing ternary hydrides in the Na-Al-H system. Moreover, experimentally established compounds such as NaAlH$_4$ and Na$_3$AlH$_6$ define the thermodynamic ground states in the Na–Al–H system, placing the considered NaAlH$_3$ composition above the ternary convex hull \cite{QiuOpalkaOlsonAnton}.

\begin{table}[h]
\begin{ruledtabular}
    \caption{Lattice constant $a_0$, volume $V$, and bond lengths Al--H and Na--H in NaAlH$_3$ under low pressure.}
    \centering
    \renewcommand{\arraystretch}{1.2}
    \begin{tabular}{l| c c c c}
        Pressure (GPa) & $a_0$ (\AA) & $V$ (\AA$^3$) & Al--H (\AA) & Na--H (\AA) \\\hline
        0.0 & 3.789 & 54.382 & 1.894 & 2.679 \\
        0.1 & 3.786 & 54.254 & 1.893 & 2.677 \\
        0.2 & 3.783 & 54.127 & 1.891 & 2.675 \\
    \end{tabular}
    \label{tab01}
\end{ruledtabular}
\end{table}

\begin{figure}[!h]
    \centering
    \includegraphics[width=1\linewidth]{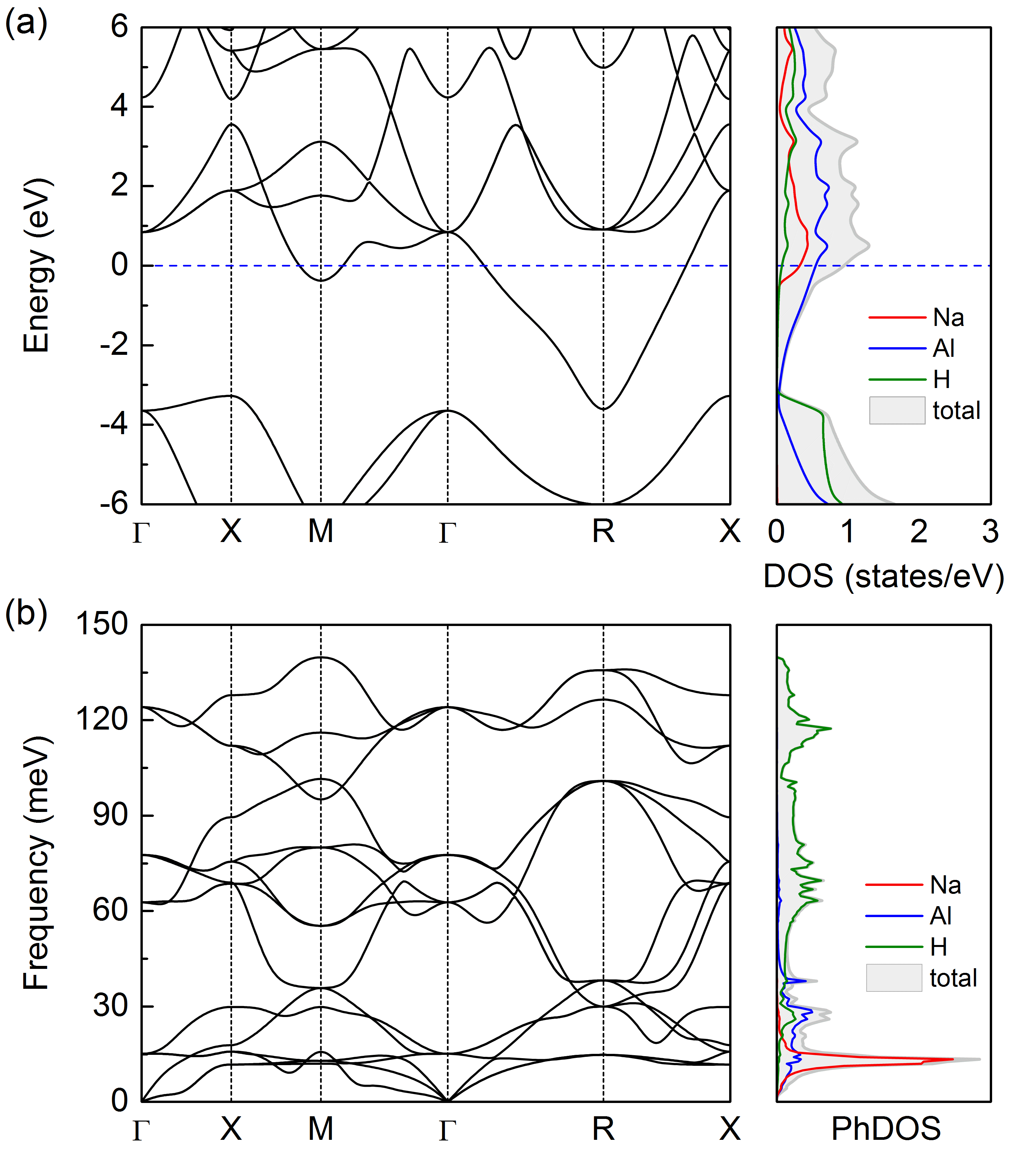}
    \caption{Electronic band structures together with partial density of states (a) and phonon dispersion together with projected phonon density of states (b) for NaAlH$_3$ at ambient pressure. The blue dashed line in (a) indicates the Fermi level.
    }
    \label{fig02}
\end{figure}

To further investigate the electronic and vibrational properties of cubic NaAlH$_3$, we computed its electronic band structure, atom-projected density of states (DOS), phonon dispersion, and phonon density of states (PhDOS), as presented in \fig{fig02}. The electronic band structure, evaluated along the high-symmetry path $\Gamma$--X--M--$\Gamma$--R--X in the Brillouin zone, demonstrates a metallic character, with a dispersive conduction band crossing the Fermi level (indicated by the dashed blue line). This feature implies the presence of partially filled electronic states and delocalized carriers, which are favorable for phonon-mediated superconductivity.

The atom-projected DOS reveals that Al-derived states (blue line) dominate in the vicinity of the Fermi level, with significant contributions from Na (red line) as well, while H-derived states (green line) are mainly located in the lower valence region, particularly around -4~eV. The finite total DOS at the Fermi level arises mainly from Al and Na states, consistent with the metallic character of the system. The presence of H contributions deeper in the valence band suggests strong Al–H bonding. This electronic structure reflects hybridized bonding characteristics and is consistent with a covalent–metallic environment, which can support enhanced electron–phonon coupling within the present theoretical framework~\cite{Gao2021, Sanna2024}.

The phonon dispersion relations presented in \fig{fig02}(b) show no imaginary frequencies throughout the Brillouin zone within the adopted computational framework, indicating dynamical stability of the considered $Pm\overline{3}m$ phase of NaAlH$_3$ at ambient pressure. The corresponding projected phonon density of states (PhDOS) provides further insight into the vibrational characteristics of the system. The low-frequency acoustic modes are dominated by Na atomic motion, while Al atoms contribute broadly across the intermediate-frequency range. In contrast, the high-frequency optical modes are primarily associated with H vibrations. This separation of phonon branches reflects the mass disparity among the constituent elements and is consistent with strong Al–H bonding. The phonon dispersion relations of NaAlH$_3$ calculated at selected pressures between $0$ and $5$~GPa are provided in the Supplemental Material~\cite{SM}.

\begin{figure}[!h]
    \centering
    \includegraphics[width=1\linewidth]{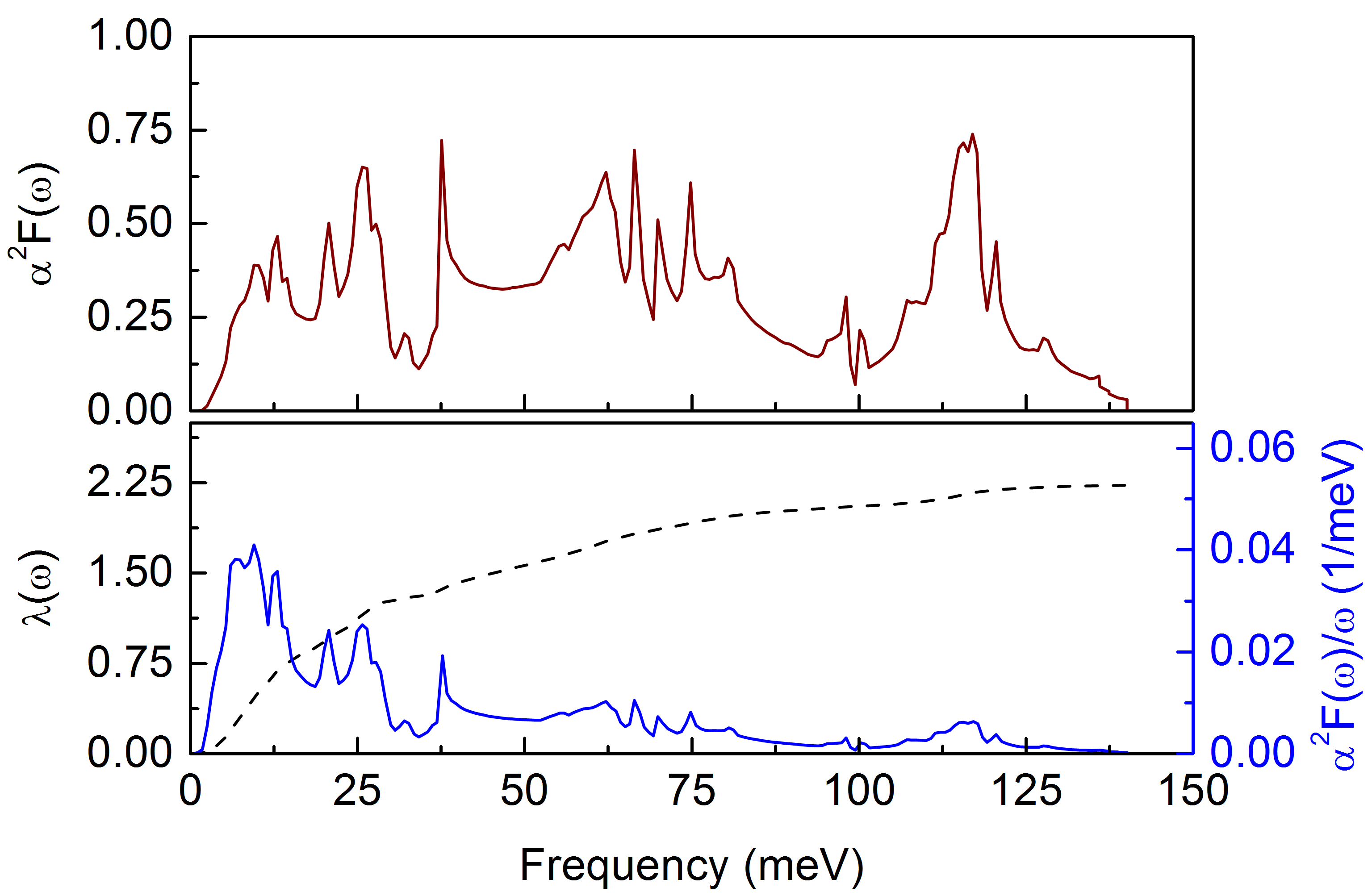}
    \caption{Isotropic Eliashberg spectral function $\alpha^2F(\omega)$ (top panel) and cumulative electron-phonon coupling parameter $\lambda(\omega)$, and its derivative with respect to frequency ${d\lambda(\omega)}/{d\omega}={\alpha^2F(\omega)}/{\omega}$ (bottom panel) for $Pm\overline{3}m$-NaAlH$_3$ at ambient pressure.}
    \label{fig03}
\end{figure}

Figure~\ref{fig03} presents the calculated isotropic Eliashberg spectral function $\alpha^2F(\omega)$ (top panel) and the cumulative electron-phonon coupling (EPC) parameter $\lambda(\omega)$, alongside its frequency derivative  $d\lambda(\omega)/d\omega$, which corresponds to $\alpha^2F(\omega)/\omega$ (bottom panel). The spectral function $\alpha^2F(\omega)$ exhibits several prominent peaks across the phonon spectrum, reflecting vibrational modes that couple strongly to electronic states near the Fermi level. The high-frequency region above 100 meV is dominated by hydrogen vibrations, consistent with the light mass of H atoms. In contrast, the peaks below 50 meV are primarily associated with vibrations of the heavier Na and Al atoms.
The bottom panel of \fig{fig03} shows the progressive accumulation of electron-phonon coupling via $\lambda(\omega)$, which reaches a total value of $\lambda = 2.23$. This large EPC constant signals very strong electron-phonon interactions, consistent with the possibility of high-temperature phonon-mediated superconductivity. The function $\lambda(\omega)$ exhibits a steep rise at low frequencies, particularly below $30$ meV, indicating that low-energy phonon modes, mainly involving Na and Al, contribute significantly to the total coupling strength.
The derivative of $\lambda(\omega)$ with respect to frequency, shown as a solid blue line, corresponds to $\alpha^2F(\omega)/\omega$ and highlights the frequency-resolved contribution to the electron-phonon coupling. Regions where this function attains large values directly correlate with rapid increases in $\lambda(\omega)$, thereby identifying the phonon modes most effective in mediating electron pairing. The pronounced peak of $\alpha^2F(\omega)/\omega$ in the low-frequency range emphasizes the dominant role of low-energy phonons, primarily associated with Na vibrations, in driving the steep initial increase of $\lambda(\omega)$.

Analytical formulas such as the McMillan \cite{McMillan1968} and Allen-Dynes \cite{allen1975A, allen1975B} equations are commonly used to estimate the superconducting critical temperature ($T_c$), but they rely on simplifying assumptions, including weak or moderate electron-phonon coupling strength ($\lambda \lesssim 1.5$). As a result, their accuracy diminishes in the strong-coupling regime. To achieve a more reliable description of the superconducting state in such systems, we employed the Migdal-Eliashberg formalism, which provides a more rigorous treatment of electron-phonon interactions beyond the Bardeen-Cooper-Schrieffer (BCS) approximation. Figure~\ref{fig04} presents the temperature evolution of two key superconducting observables: the superconducting energy gap $\Delta(T)$ and the specific heat difference between the superconducting and normal states $\Delta C(T) = C_s(T) - C_n(T)$.
In particular, \fig{fig04}(a) shows that the energy gap $\Delta(T)$ decreases smoothly with increasing temperature and vanishes continuously at the critical temperature $T_c$, following a behavior consistent with second-order phase transitions. The magnitude of the zero-temperature gap $\Delta(0)$, as well as the critical temperature $T_c$, depends sensitively on the choice of the Coulomb pseudopotential $\mu^\ast$. For lower values of $\mu^\ast$, a stronger pairing interaction is observed, leading to larger energy gaps and higher $T_c$ values. Specifically, at ambient pressure, $\Delta(0)$ ranges from $15.46$ meV to $12.65$ meV for $\mu^\ast$ varying between $0.1$ and $0.2$, respectively.

\begin{figure}[!t]
    \centering
    \includegraphics[width=1\linewidth]{Fig04}
    \caption{Superconducting energy gap as a function of temperature (a) and temperature-dependent specific heat difference between the superconducting and normal state (b) for $Pm\overline{3}m$-NaAlH$_3$ at ambient pressure for three values of Coulomb pseudopotential $\mu^{\ast}=0.1$ (red lines), $\mu^{\ast}=0.15$ (blue lines), and $\mu^{\ast}=0.2$ (green lines). 
    }
    \label{fig04}
\end{figure}

Figure~\ref{fig04}(b) illustrates the temperature dependence of the specific heat difference. A clear discontinuity appears at $T_c$ for each case, marking the transition between the superconducting and normal states. The magnitude of the specific heat jump $\Delta C(T_c)$ increases with decreasing $\mu^\ast$, indicating stronger superconducting correlations. Notably, the calculated values of $\Delta C(T_c)/\gamma T_c$ significantly exceed the BCS weak-coupling limit of $1.43$ \cite{BCS}, reaching values as high as $2.28$ (for $p=0$ GPa and $\mu^{^\ast}=0.2$), which confirms the strong-coupling nature of the superconductivity in NaAlH$_3$.

\begin{table}[h]
\begin{ruledtabular}
    \caption{Superconducting properties of NaAlH$_3$ as a function of pressure $p\in\{0, 0.1, 0.2\}$ GPa and Coulomb pseudopotential $\mu^{\star}\in\{0.1, 0.15, 0.2\}$.}
    \centering
    \renewcommand{\arraystretch}{1.2}
    \begin{tabular}{l| l | c c c}
 
        \multirow{2}{*}{\textbf{}} & \multirow{2}{*}{$\mu^{\star}$} & \multicolumn{3}{c}{pressure (GPa)} \\
        \cmidrule(lr){3-5}      
        & & 0 & 0.1 & 0.2 \\\hline
        \midrule
        $\lambda$ & - & 2.230 & 2.225 & 2.223 \\
        N(E$_F$) (states/eV) & - & 0.942 & 0.942 & 0.942 \\
        $\gamma$ (mJ/mol$\cdot$K$^2$) & - & 14.345 & 14.340 & 14.338 \\\hline 
        \midrule
        \multirow{3}{*}{$T_c$ (K)} & 0.1 & 73.7 & 73.3 & 73.2  \\
        & 0.15 & 66.7 & 66.5 & 66.0 \\
        & 0.2 & 61.4 & 61.2 & 61.1 \\\hline 
        \midrule
        \multirow{3}{*}{$\alpha$} & 0.1 & 0.427 & 0.437 & 0.441 \\
        & 0.15 & 0.420 & 0.430 & 0.437 \\
        & 0.2 & 0.415 & 0.426 & 0.427 \\\hline 
        \midrule
        \multirow{3}{*}{$\Delta(0)$ (meV)} & 0.1 & 15.462 & 15.339 & 15.203 \\
        & 0.15 & 13.857 & 13.734 & 13.598 \\
        & 0.2 & 12.647 & 12.524 & 12.388 \\\hline 
        \midrule
        \multirow{3}{*}{$2\Delta(0)/k_BT_c$} & 0.1 & 4.869 & 4.857 & 4.820 \\
        & 0.15 & 4.822 & 4.793 & 4.782 \\
        & 0.2 & 4.781 & 4.750 & 4.706 \\\hline 
        \midrule
        \multirow{3}{*}{$\Delta C(T_c)$ (J/mol$\cdot$K)} & 0.1 & 2.359 & 2.304 & 2.287 \\
        & 0.15 & 2.157 & 2.141 & 2.128 \\
        & 0.2 & 2.011 & 2.004 & 1.998 \\\hline 
        \midrule
        \multirow{3}{*}{$\Delta C(T_c)/\gamma T_c$} & 0.1 & 2.231 & 2.192 & 2.179 \\
        & 0.15 & 2.255 & 2.245 & 2.249 \\
        & 0.2 & 2.284 & 2.283 & 2.281 \\
        \bottomrule
    \end{tabular}
    \label{tab:superconductivity}
    \end{ruledtabular}
\end{table}

Complementary quantitative data are presented in Table~\ref{tab:superconductivity}, summarizing key superconducting parameters of NaAlH$_3$ for three representative values of the Coulomb pseudopotential $\mu^\ast$ (0.1, 0.15, and 0.2) under hydrostatic pressures of 0, 0.1, and 0.2 GPa. The electron-phonon coupling constant remains remarkably high ($\lambda = 2.23$) and exhibits minimal sensitivity to pressure, underscoring the robustness of the coupling mechanism. While a slight decrease in the electronic density of states at the Fermi level $N(E_F)$ and the Sommerfeld coefficient $\gamma$ is observed with increasing pressure, these variations do not significantly affect the superconducting performance.

Indeed, the critical temperature $T_c$ shows a mild increase with pressure across all values of $\mu^\ast$, reflecting subtle modifications in the phonon spectrum and electronic structure. Similarly, the superconducting energy gap $\Delta(0)$ and the dimensionless gap ratio $2\Delta(0)/k_B T_c$ also increase slightly, with the latter remaining well above the BCS limit of 3.53~\cite{BCS}, consistent with strong-coupling superconductivity. Notably, NaAlH$_3$ achieves a high $T_c$ of $\sim$74 K at ambient pressure (for $\mu^\ast = 0.1$), surpassing its isostructural analogs KAlH$_3$ and RbAlH$_3$, which require external pressures of 0–7 GPa and 15 GPa, respectively, to reach comparable critical temperatures~\cite{Mingyang_2024, Wines_2024}. This superior performance is attributed in part to its stronger electron-phonon interaction, as evidenced by the higher $\lambda$ compared to the K/Rb counterparts ($\lambda \sim 1.8$–2.2).

We further investigate the nature of superconductivity in NaAlH$_3$ by analyzing the isotope effect on the critical temperature. The isotope coefficient ($\alpha$) is defined as: 
\begin{equation} \label{alpha} \alpha = -\frac{{\rm ln}[T_{c}]_{{\rm NaAlD}{_3}} - {\rm ln}[T_{c}]_{{\rm NaAlH}{_3}}}{{\rm ln}[M]_{\rm D} - {\rm ln}[M]_{\rm H}}, 
\end{equation} 
where $[M]_{\rm H}$ and $[M]_{\rm D}$ are the atomic masses of hydrogen and deuterium, respectively. Within the framework of conventional BCS theory, a weak-coupling superconductor exhibits an isotope coefficient of $\alpha = 0.5$. In contrast, strong electron-phonon coupling and retardation effects, as captured by the Migdal-Eliashberg theory, typically reduce the $\alpha$ value below $0.5$. For example, $\alpha$ values of $0.353$ (H$_3$S) and $0.465$ (LaH$_{10}$) have been predicted using the Migdal-Eliashberg theory \cite{Durajski_2016_isotope, Durajski2020}.
The phonon dispersion relations of NaAlH$_3$ and NaAlD$_3$, calculated at $0$ GPa, $0.1$ GPa, and $0.2$ GPa, are presented in SM \cite{SM}. Our calculations show that the isotope coefficient for NaAlH$_3$ falls between $0.415$ and $0.441$ (see Table~\ref{tab:superconductivity}). While slightly below the ideal BCS value of $0.5$, these values remain comparatively high and, together with the large electron-phonon coupling constant, provide strong evidence that superconductivity in NaAlH$_3$ is driven by electron-phonon interactions.

\begin{figure}[!t]
    \centering
    \includegraphics[width=1\linewidth]{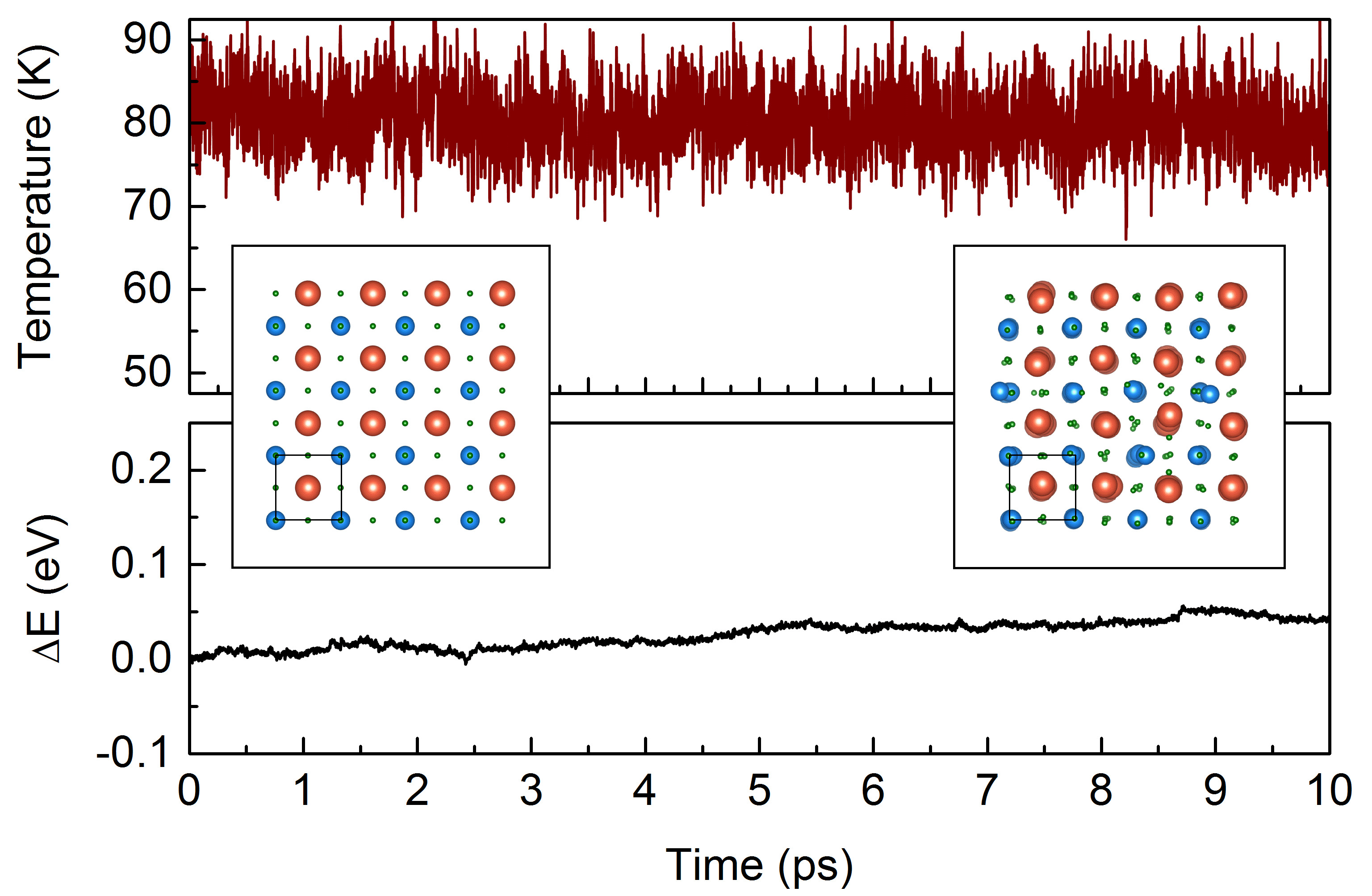}
    \caption{Fluctuations of temperature (top panel) and total energy difference $\Delta E$ (bottom panel) during
AIMD simulations at $80$ K. 
    }
    \label{fig05}
\end{figure}

To assess the finite-temperature behavior of the hypothetical NaAlH$_3$ phase slightly above $T_c$, we performed \textit{ab initio} molecular dynamics (AIMD) simulations in the canonical (NVT) ensemble. The simulations were carried out using a $4\times4\times4$ supercell containing $320$ atoms at $80$ K, with a time step of $1$ fs over a total simulation time of $10$ ps. As shown in \fig{fig05}, both the total energy difference ($\Delta E = E - E(0)$) and the temperature exhibit only small fluctuations around their equilibrium values, and no significant structural distortions or bond-breaking events are observed during the simulated time window. These results indicate that the considered cubic $Pm\overline{3}m$ structure remains dynamically intact under the applied simulation conditions, within the limitations of the AIMD time scale.

\section{Summary}

In summary, we have investigated the electronic structure, lattice dynamics, and superconducting properties of a hypothetical cubic $Pm\overline{3}m$ phase of NaAlH$_3$ using first-principles calculations combined with the Migdal–Eliashberg formalism. Within the adopted theoretical framework, this phase exhibits exceptionally strong electron–phonon coupling ($\lambda = 2.23$), leading to a high superconducting critical temperature of up to approximately $74$ K for a reasonable choice of the Coulomb pseudopotential.
The calculated superconducting gap ratio and specific heat jump substantially exceed the BCS weak-coupling limits, highlighting the strong-coupling nature of superconductivity in this system. The underlying mechanism is associated with the interplay between Al–H covalent bonding and Na-derived metallic states, which gives rise to favorable electronic and vibrational characteristics. We emphasize that the considered NaAlH$_3$ structure is thermodynamically metastable with respect to competing ternary hydrides, and its predicted properties should therefore be regarded as those of a theoretical model system, rather than an experimentally established compound.
Despite these limitations, the present results provide insight into how alanate-based frameworks can host strong electron–phonon coupling at ambient pressure, and may inform future theoretical searches for experimentally accessible ternary hydrides with enhanced superconducting properties.


\section{Competing interest}
The authors declare no competing financial interest.

\section{Data availability}
The data containing information needed to reproduce presented results are available in the public database of Zenodo under accession code 17338327 \cite{zenodo}.

\section{Acknowledgements}
A.P. Durajski acknowledges financial support from the National Science Centre (Poland) under Project No. 2022/47/B/ST3/00622. Y. Li acknowledges the funding from the NSFC under Grants No. 12474012 and No. 12074154, and the funding from the 333 High-level Talents Project of Jiangsu Province.
We gratefully acknowledge Polish high-performance computing infrastructure PLGrid (HPC Center: ACK Cyfronet AGH) for providing computer facilities and support within computational grant no. PLG/2025/018759.


\section{References}
\bibliography{bibliography}

\end{document}